\documentclass[prb,superscriptaddress,twocolumn,floatfix,amsmath,amssymb]{revtex4}
\usepackage{graphicx}
\usepackage{dcolumn}
\usepackage{verbatim}
\usepackage{times}
\usepackage{caption}
\usepackage{subfigure}
\usepackage{bm}
\usepackage{color}
\usepackage[colorlinks,bookmarks=false,citecolor=blue,linkcolor=red,urlcolor=blue,backref=red,dvipdfm]{hyperref}
\usepackage{booktabs}
\usepackage{appendix}
\begin{document}

\title{Node-line Dirac semimetal manipulated by Kondo mechanism in nonsymmorphic CePt$_2$Si$_2$}

\author{Hao-Tian Ma}
\affiliation{College of Science, Guilin University of Technology, Guilin 541004, China}
\author{Xing Ming}
\affiliation{College of Science, Guilin University of Technology, Guilin 541004, China}
\author{Xiao-Jun Zheng}
\affiliation{College of Science, Guilin University of Technology, Guilin 541004, China}
\author{Jian-Feng Wen}
\affiliation{College of Science, Guilin University of Technology, Guilin 541004, China}
\author{Yue-Chao Wang}
\affiliation{Laboratory of Computational Physics, Institute of Applied Physics and Computational Mathematics, Beijing 100088, China}
\author{Yu Liu}
\affiliation{Laboratory of Computational Physics, Institute of Applied Physics and Computational Mathematics, Beijing 100088, China}
\author{Huan Li}
\email{lihuan@glut.edu.cn}
\affiliation{College of Science, Guilin University of Technology, Guilin 541004, China}

\date{\today}

\begin{abstract}

Dirac node lines (DNLs) are characterized by Dirac-type linear crossings between valence and conduction bands along one-dimensional node lines in the Brillouin zone (BZ). Spin-orbit coupling (SOC) usually shifts the degeneracy at the crossings thus destroys DNLs, and so far the reported DNLs in a few materials are non-interacting type, making the search for robust interacting DNLs in real materials appealing.
Here, via first-principle calculations, we reveal that Kondo interaction together with nonsymmorphic lattice symmetries can drive a robust interacting DNLs in a Kondo semimetal CePt$_2$Si$_2$, and the feature of DNLs can be significantly manipulated by Kondo behavior in different temperature regions.
Based on the density function theory combining dynamical mean-field theory (DFT+DMFT), we predict a transition to Kondo-coherent state at coherent temperature $T_{\mathrm{coh}}\approx$ 80 K upon cooling, verified by temperature dependence of Ce-4$f$ self-energy, Kondo resonance peak, magnetic susceptibility and momentum-resolved spectral. Below $T_{\mathrm{coh}}$, well-resolved narrow heavy-fermion bands emerge near the Fermi level, constructing clearly visualized interacting DNLs locating at the BZ boundary, in which the Dirac fermions have strongly enhanced effective mass and reduced velocity.
In contrast, above a crossover temperature $T_{\mathrm{KS}}\approx$ 600 K, the destruction of local Kondo screening drives non-interacting DNLs which are comprised by light conduction electrons at the same location. These DNLs are protected by lattice nonsymmorphic symmetries thus robust under intrinsic strong SOC.
Our proposal of DNLs which can be significantly manipulated according to Kondo behavior provides an unique realization of interacting Dirac semimetals in real strongly correlated materials, and serves as a convenient platform to investigate the effect of electronic correlations on topological materials.

\end{abstract}

\maketitle

\section{Introduction}

Dirac semimetals, such as Na$_3$Bi~\cite{Wang12} and Cd$_3$As$_2$~\cite{Wang13}, are characterized by linear crossings between valence and conduction bands in momentum space, forming four-fold-degenerate Dirac points describing by Dirac equation, and can be viewed as three-dimensional (3D) analogy of the two-dimensional (2D) Dirac points in graphene. Under breaking of either time-reversal or space-inversion symmetry, individual Dirac point can be divided into a pair of Weyl points with opposite chiralities, as observed in TaAs~\cite{Lv15} and Ag$_2$S~\cite{Wang19}, etc. In some materials with negligible spin-orbital coupling (SOC), the valence and conduction bands meet along a curved line or closed loop in the Brillouin zone, forming Dirac node lines (DNLs), which are usually unstable under action of SOC. Recently, DNLs in materials with space groups No. 129 and 125 are proposed and confirmed by angle-resolved
photoemission spectrums (ARPES) observations. In these materials such as ZrSiS~\cite{Fu17,Schoop15,Chen17} and PtPb$_4$~\cite{Wu22}, the DNLs are protected by nonsymmorphic symmetries in their lattice space group, making them robust under SOC. Additionally, the Dirac fermions on these DNLs exhibit 2D character in momentum space~\cite{Young15}.

In contrast to ZrSiS and PtPb$_4$, in which the electronic correlations are negligible, strong correlations may bring dramatic affects to the DNLs. As in the case of Weyl-Kondo semimetal Ce$_3$Bi$_4$Pt$_3$~\cite{Lai18,Cao20}, the strong correlation and Kondo hybridization result in renormalized Weyl fermions with highly enhanced effective mass and suppressed Fermi velocity, leading to characteristic $T^3$ dependence of specific heat~\cite{Lai18}, more remarkably, the correlations can give rise to nonlinear response behaviors such as giant spontaneous Hall effect~\cite{Kofuji21,Dzsaber21}. Similarly, the electron correlations in material with DNLs may also induce notable affects to the Dirac fermions and arouse anomalous transport phenomena which can be explored in future experiments, nevertheless, such interacting DNLs seem lack of report in the literature yet.

In this article, we systematically explore nonsymmorphic Kondo semimetals CePt$_2$Si$_2$ and CePt$_2$Ge$_2$ by density-function theory combing with dynamical mean-field theory (DFT+DMFT). Firstly, we find that below a characteristic Kondo coherence temperature at about 80 K, CePt$_2$Si$_2$ becomes Kondo-coherent, forming interacting DNLs by heavy quasi-particles composed of conduction $spd$ electrons and Ce-$4f$ electrons. Secondly, above another crossover temperature at about 600 K, the local Kondo screening in CePt$_2$Si$_2$ has been destructed, hence the DNLs become non-interacting, consisting of only light conduction electrons. Due to larger unit cell volume, the DNLs in CePt$_2$Ge$_2$ remains non-interacting in all calculated temperature region.
DNLs are also reported in a few rare-earth compounds such as centrosymmetric CeRhSb and CeNiSn~\cite{Nam19}, however, the correlated 4$f$ electrons in those materials are already localized thus their DNLs are actually created by conduction electrons only. In this context, to the best of our knowledge, CePt$_2$Si$_2$ provides the first studied material with interacting DNLs in real strongly correlated materials.

The rest of this paper is arranged as follows. In section 2, we will analyse the crystal structures of CePt$_2$Si$_2$ and CePt$_2$Ge$_2$, and introduce the technical details of DFT+DMFT simulations in present work. In section 3, we will present the DFT+DMFT results of CePt$_2$Si$_2$. Through synthetically analysis of self-energy, density of state, momentum-resolved spectral and the magnetic susceptibility, we will evaluate the two characteristic temperature, firstly the Kondo coherent temperature driving the formation of coherent heavy-fermion hybridization bands, secondly the Kondo screening temperature below which the local Kondo screening of $4f$ electrons by conduction electrons turns on.
In sections 4, we will discuss the emergence of interacting DNLs in CePt$_2$Si$_2$ below Kondo-coherent temperature, and also the non-interacting DNLs above the Kondo screening temperature.
We will also verify the non-interacting DNLs in CePt$_2$Ge$_2$ based on its DFT and DFT+DMFT results.
The last section will give a brief conclusion and discussion.

\section{crystal structure and computational method}

\begin{figure}[tbp]
\hspace{-0cm} \includegraphics[totalheight=1.5in]{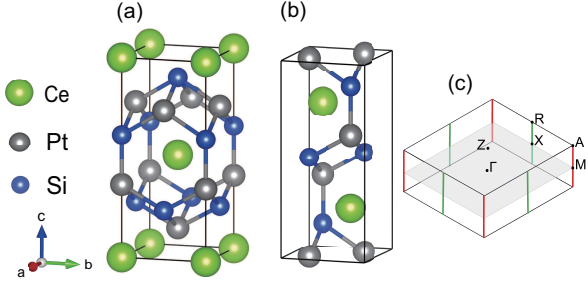}
\caption{
(a) Primitive unit cell of CePt$_2$Si$_2$ with tetragonal CaBa$_2$Ge$_2$ type structure. (b) Equivalent unit cell displaying inversion and nonsymmorphic symmetries. (c) Corresponding Brillouin zone, in which the red and green lines denotes the DNLs along M-A and X-R paths, respectively.
}
\label{lattice}
\end{figure}

The primitive unit cell of CePt$_2$Si$_2$ is illustrated in Fig.\ref{lattice}(a), which crystalizes
in the tetragonal CaBa$_2$Ge$_2$ type structure with space group $P4/nmm$ (No. 129). The lattice constants and atom positions of CePt$_2$Si$_2$ and isostructural CePt$_2$Ge$_2$ are collected in Tab.\ref{tab1} according to Ref.~\onlinecite{Dommann85}. From an alternate set of unit cell of CePt$_2$Si$_2$ in Fig.\ref{lattice}(b), the space inversion symmetry can be clearly seen. Besides the inversion symmetry, CePt$_2$Si$_2$ and CePt$_2$Ge$_2$ exhibit nonsymmorphic symmetries combining point group and fractional translation operations, namely the gliding mirror plane $\{M_z|\frac{1}{2},\frac{1}{2}\}$ and screw axies $\{C_{2x}|\frac{1}{2},0\}$, $\{C_{2y}|0,\frac{1}{2}\}$, where the origin of axies locate at the center of the unit cell in Fig.\ref{lattice}(b). CePt$_2$Si$_2$ was found to remain paramagnetic as low as 0.06 K~\cite{Reotier97}, signaling the preservation of time-reversal symmetry.
The nonsymmorphic symmetries combining time-reversal and inversion symmetries protect the four-fold degeneracy at X and M, and also along X-R and M-A lines in the Brillouin zone, creating the DNLs shown in Fig.\ref{lattice}(c), as will be discussed in detail below.

\heavyrulewidth=1bp

\begin{table*}
\small
\renewcommand\arraystretch{1.3}
%\centering
\caption{\label{tab1}
Crystal parameters of CePt$_2$Si$_2$ and CePt$_2$Ge$_2$~\cite{Dommann85}, the atomic positions correspond to Fig.\ref{lattice}(b). }
%\begin{ruledtabular}
\begin{tabular*}{17cm}{@{\extracolsep{\fill}}ccccccccc}
\toprule
        & lattice parameters          & Ce position & Pt position & Si position  \\
\hline
 CePt$_2$Si$_2$ & a=b=4.252{\AA} & Ce(2c):($\frac{1}{4}$,$\frac{1}{4}$,0.7452),($\frac{3}{4}$,$\frac{3}{4}$,0.2548) & Pt(2c):($\frac{1}{4}$,$\frac{1}{4}$,0.3798),($\frac{3}{4}$,$\frac{3}{4}$,0.6202)  & Si(2c):($\frac{1}{4}$,$\frac{1}{4}$,0.1329),($\frac{3}{4}$,$\frac{3}{4}$,0.8671)  \\
      &  c=9.788{\AA} &  & Pt(2a):($\frac{1}{4}$,$\frac{3}{4}$,0),($\frac{3}{4}$,$\frac{1}{4}$,0) &          Si(2b):($\frac{1}{4}$,$\frac{3}{4}$,$\frac{1}{4})$,($\frac{3}{4}$,$\frac{1}{4}$,$\frac{1}{2}$)
\\
 CePt$_2$Ge$_2$ & a=b=4.397{\AA} & Ce(2c):($\frac{1}{4}$,$\frac{1}{4}$,0.74),($\frac{3}{4}$,$\frac{3}{4}$,0.26) & Pt(2c):($\frac{1}{4}$,$\frac{1}{4}$,0.383),($\frac{3}{4}$,$\frac{3}{4}$,0.617)  & Ge(2c):($\frac{1}{4}$,$\frac{1}{4}$,0.131),($\frac{3}{4}$,$\frac{3}{4}$,0.869)  \\
      &  c=9.802{\AA} &  & Pt(2a):($\frac{1}{4}$,$\frac{3}{4}$,0),($\frac{3}{4}$,$\frac{1}{4}$,0) &          Ge(2b):($\frac{1}{4}$,$\frac{3}{4}$,$\frac{1}{4})$,($\frac{3}{4}$,$\frac{1}{4}$,$\frac{1}{2}$)\\
\bottomrule
\end{tabular*}
%\end{ruledtabular}
\label{data}
\end{table*}

In order to explore the electron-correlation effects in CePt$_2$Si$_2$ and CePt$_2$Ge$_2$, we employ the density function theory combined with single-site dynamical mean-field theory (DFT+DMFT) embodied in the EDMFT package~\cite{Haule10}, with the DFT part implemented by full-potential linear augmented plane-wave method built in WIEN2k code~\cite{Blaha20}. Such DFT+DMFT method has been successfully applied in studying electronic correlations in a variety of materials, especially in rear-earth compounds~\cite{Nam19,Lu16,Shim07,Zhu20,Wang21,Chen18}. In the DFT part, we use a $16\times16\times7$ k-mesh in the Brillouin zone integration, with a cut-off parameter $K_{\mathrm{max}}$ given by $R_{\mathrm{MT}} K_{\mathrm{max}}=7.0$, and spin-orbital coupling (SOC) is included throughout the calculations. The DFT band structures are also cross-checked by VASP code. Each DFT+DMFT step contains one-shot DMFT and 20 steps of DFT calculations. In the DMFT iterations, in order to better fit experiment observations, we employ the on-site Coulomb repulsion $U=5.0$ eV and Hund's coupling $J_H=0.76$ eV on Ce-4f orbits, similar to the value set in Refs.~\onlinecite{Shim07} and ~\onlinecite{Nam19}. The states within energy window [-10 eV,10 eV] from the Fermi level are projected into the Anderson impurity problems.
We use the continuous-time quantum Monte Carlo method (CT-QMC) to solve the Anderson impurity problems, then perform analytical continuation by maximum-entropy method to obtain the real-frequency self-energies for $f$ electrons. In each CT-QMC computation, 128 CPU cores are used to run (5$\sim$30)$\times 10^8$ QMC steps from 1000 K to 10 K. Typically, within 30$\sim$40 DFT+DMFT iterations, full-charge self consistence can be reached, then we run additional 5 iterations to further average the self-energies. Since no magnetic order was found in CePt$_2$Si$_2$ and CePt$_2$Ge$_2$ down to 60 mK~\cite{Reotier97}, we focus on the paramagnetic phase.
In the impurity solver, the crystal-field splitting of $f$ orbits has been examined and is found to be one more orders of magnitude smaller than the SOC splitting, hence the crystal-field splitting is neglected in present calculations. The SOC splits Ce-4$f$ orbits into $j=5/2$ and $j=7/2$ states, denoted by $4f_{5/2}$ and $4f_{7/2}$ respectively in the following.

\section{correlation effect and Kondo behavior}

\begin{figure}[tbp]
\hspace{-0cm} \includegraphics[totalheight=1.6in]{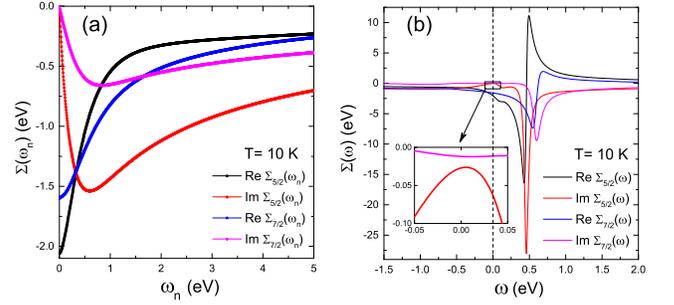}
\caption{Self-energies of $4f_{5/2}$ and $4f_{7/2}$ states via DFT+DMFT calculations of CePt$_2$Si$_2$ at 10 K on (a) imaginary-frequency axis and (b) real-frequency axis. Inset of (b) shows the detail near zero frequency.
}
\label{Selfenergy}
\end{figure}

In DFT+DMFT formulations, the electron correlation manifests itself following the local self-energies of $f$ states. In Fig.\ref{Selfenergy}, we plot the calculated DFT+DMFT local self-energies $\Sigma(\omega_n)$ and $\Sigma(\omega)$ for CePt$_2$Si$_2$ at temperature $T$=10 K, on imaginary- and real-frequency axes respectively, where Matsubara frequency $\omega_n=(2n+1)\pi T$. In Fig.\ref{Selfenergy} (a), the imaginary-part self-energies for both $4f_{5/2}$ and $4f_{7/2}$ states nicely approach zero at zero imaginary frequency, indicating Fermi-liquid like character.
The slope of imaginary-part self-energy $\mathrm{Im}\Sigma_{5/2}(\omega_n)$ for $4f_{5/2}$ state at zero frequency is about -9.3, which gives the quasi-particle spectral weight and mass enhancement factor
\begin{align}
Z=1/(1-\frac{\partial\Sigma(\omega_n)}{\partial\omega_n}|_{\omega_n\rightarrow0^+})=0.097, \nonumber\\
m^{\ast}/m_{DFT}=1/Z=10.3, \label{Z}
\end{align}
respectively, it means that the band width of $4f_{5/2}$ bands are strongly reduced to be roughly ten times narrow than in DFT, as will be verified in the following. Large value of $m^{\ast}/m_{DFT}$ is the origin of large specific heat of CePt$_2$Si$_2$ observed at low temperature~\cite{Ayache87}.
The self-energies on real axis are created through analytical continuation of $\Sigma(\omega_n)$ by maximum entropy method, and are shown in Fig.\ref{Selfenergy} (b). The real-part self-energies for both $4f_{5/2}$ and $4f_{t/2}$ states show rapid variations between $\omega$=(0.2 eV, 1 eV), leading to significant modifies of $f$ bands from DFT results by correlations, as will be seen below.
At $T$=10 K, the imaginary self-energy $\mathrm{Im}\Sigma_{5/2}(\omega)$ of $4f_{5/2}$ state has a small value 27.2 meV at $\omega=0$, gives rise to low scattering rate and relatively long lift time for quasi particles at the Fermi level.

\begin{figure}[tbp]
\hspace{-0cm} \includegraphics[totalheight=1.4in]{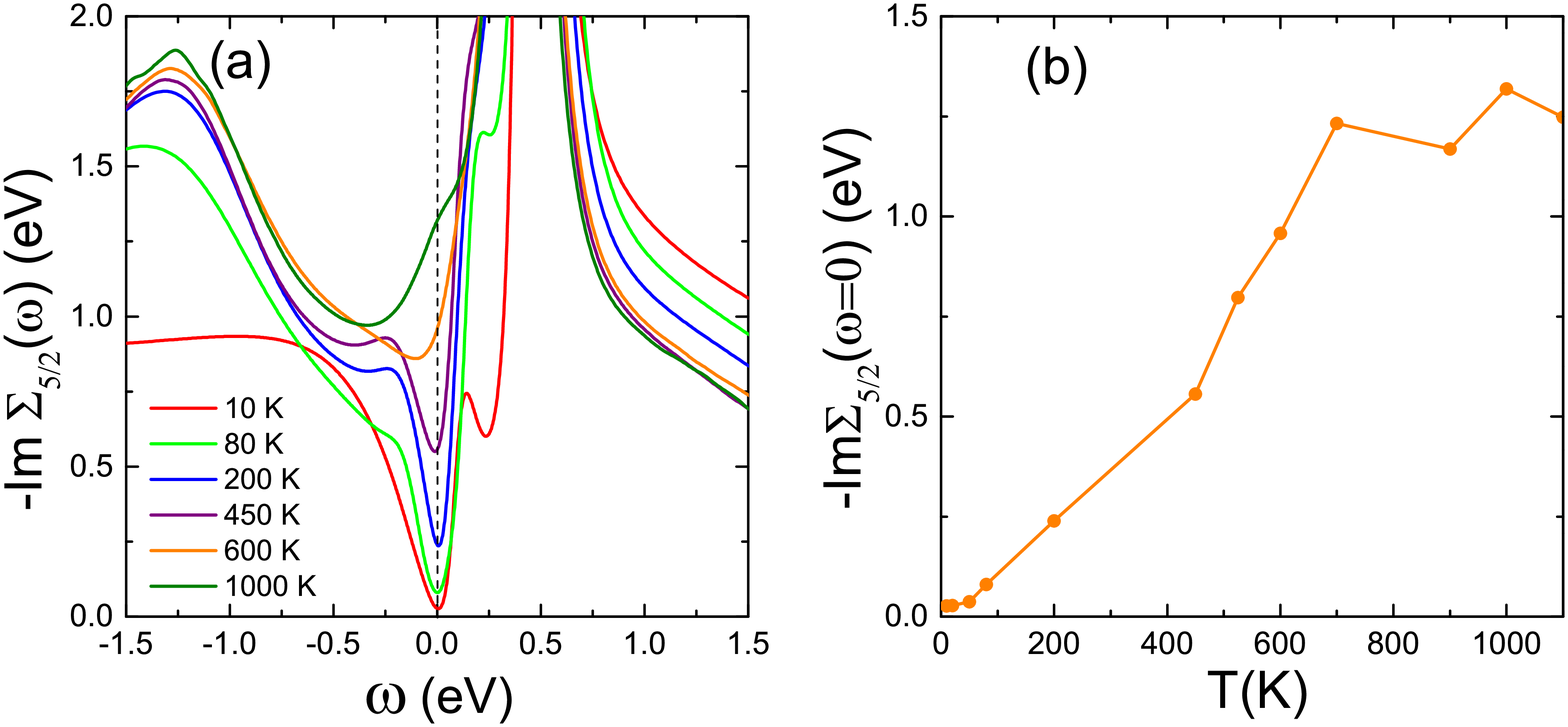}
\caption{
(a) Imaginary part of 4$f_{5/2} $self-energy Im$\Sigma_{5/2}(\omega)$ vs $\omega$ at different temperatures.
(b) Evolution of Im$\Sigma_{5/2}(\omega=0)$ with temperature.
}
\label{ImSig}
\end{figure}

Fig.\ref{ImSig}(a) demonstrates the evolution of imaginary self-energy $\mathrm{Im}\Sigma_{5/2}(\omega)$ for $4f_{5/2}$ state as a function of frequency $\omega$, at various temperatures. Below 80 K, a clear and sharp dip near $\omega=0$ appears, reaching a quite small value at the bottom. Such dip of $\mathrm{Im}\Sigma_{5/2}(\omega)$ near $\omega=0$ directly drives an intense Kondo resonance peak in the 4$f_{5/2}$ density of states (DOS) (see below), and can be interpreted as the onset of Kondo coherence~\cite{Shim07,Zhu20} below a characteristic coherent temperature $T_{\mathrm{coh}}\approx80$ K. Moreover, the sharp dip of $\mathrm{Im}\Sigma_{5/2}(\omega)$ below $T_{\mathrm{coh}}$ also induces clearly resolved 4$f$ bands, further verifying the Kondo coherence below $T_{\mathrm{coh}}$, as will be discussed below.
Besides, below 80 K, $\mathrm{Im}\Sigma_{5/2}(\omega)$ can be well fitted by a parabolic function $\mathrm{Im}\Sigma_{5/2}(\omega)\approx-\alpha(\omega-\omega_0)^2-\Sigma_0$, with $\alpha=29.7 \mathrm{eV}^{-1}$, $\omega_0=5 \mathrm{meV}$, $\Sigma_0=26.5 \mathrm{meV}$ at 10 K, also suggesting Fermi-liquid behavior at low temperature, in accordance with the interpretation from resistivity and specific heat experiments ~\cite{Ayache87,Tchokonte05}. As temperature rises, the dip of $\mathrm{Im}\Sigma_{5/2}(\omega)$ is suppressed gradually, while above 600 K, the narrow dip seem to vanish, leaving a broad minimum considerably away from $\omega=0$, resulting in a greatly weakened 4$f$ DOS. In Fig.\ref{ImSig}(b), the magnitude of $-\mathrm{Im}\Sigma_{5/2}(\omega=0)$ is plotted as temperature varies. Below 10 K, $-\mathrm{Im}\Sigma_{5/2}(0)$ tends to be saturated and approaches about 27 meV at zero temperature, while temperature rises, it shows a gradually increase, then turns to stay around large magnitude of 1.25 eV at $T>$600 K, which makes 4$f$ electrons localized at high temperature, as will be further clarified below.

\begin{figure*}[tbp]
\hspace{-0cm} \includegraphics[totalheight=2.2in]{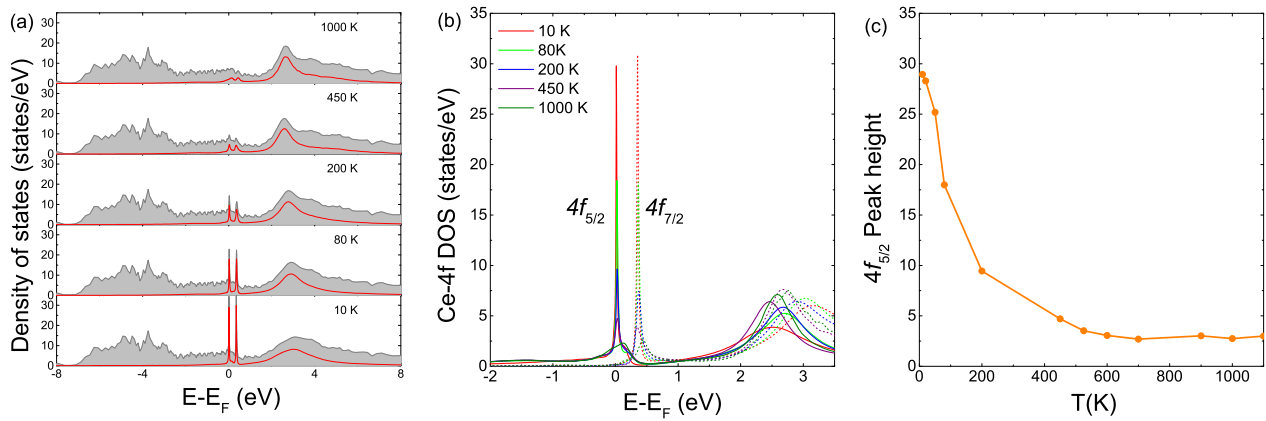}
\caption{
(a) Density of states (DOS) of CePt$_2$Si$_2$ from temperature 10 K to 1000 K via DFT+DMFT calculations. The grey solid lines denote total DOS, and the red lines denote Ce-4$f$ DOS. The paired 4$f$ peaks become more and more prominent as temperature drops. (b) Projected Ce-4$f$ DOS at different temperatures. The solid lines denote $4f_{5/2}$ DOS, while the dashed lines denote $4f_{7/2}$ DOS. The splitting of 4$f$ DOS is caused by SOC.
(c) Temperature variation of $4f_{5/2}$ peak height, implying formation of local Kondo screening and Kondo resonance below $T_{\mathrm{KS}}\approx$ 600 K.
}
\label{DOS}
\end{figure*}

The many-body Kondo screening of local $4f$ electrons by conduction electrons creates an enormous enhancement in the DOS of Ce-4$f$ states near the Fermi level, i.e., the Kondo resonance peak.
In order to further clarify the onset of Kondo coherence and Kondo screening, we calculate the DOS of CePt$_2$Si$_2$ with varying temperature. In Fig.\ref{DOS} (a), the total and 4$f$ DOS are plotted as functions of energy, from 10 K to 1000K, displaying significant difference near the Fermi level as temperature rises. At low temperatures (see 10 K and 80 K cases), two narrow peaks (peak width about 20 meV) dominated by 4$f$ states appear with large height, in which the one contributed by $4f_{5/2}$ state centers at 14 meV above the Fermi level with its tail crosses $E_F$, the other owning to $4f_{7/2}$ state locates at 0.351 eV above $E_F$, and the corresponding SOC splitting between these two peaks is about 0.337 eV, as shown in Fig.\ref{DOS} (b). As temperature increases, the two resonance peaks decrease considerably but are still visible up to 1000 K. The robust of 4$f$ DOS peaks at high temperature is owning to relatively strong impurity hybridization function in DMFT calculation, which indicates a strong $c$-$f$ hybridization in CePt$_2$Si$_2$ in wide temperature range. Since the Kondo resonance peak is dominated by the low-lying $4f_{5/2}$ state, we plot the evolution of its peak height vs temperature in Fig.\ref{DOS} (c). It can be seen that the Kondo resonance peak carries large height at low temperatures and shows a saturation tendency below 10 K, which arises from similar saturation behavior of $-\mathrm{Im}\Sigma_{5/2}(0)$ in Fig.\ref{ImSig}.
As temperature rises from 10 K, the peak height first drops rapidly, then turns to decrease much slowly, and eventually varies smoothly to maintain a small magnitude above 600 K. Around 80 K, the Kondo coherence sets in and manifests itself by a rapid increase of resonance peak, further confirming the appearance of Kondo coherence below coherent temperature $T_{\mathrm{coh}}\approx$ 80 K.
While above 600 K, Kondo resonance peak is greatly reduced, and only accounts for a small proportion in the total DOS, indicating that the 4$f$ electrons are already localized to form local moments, similar to the 4$f$ states in CeSb, CeIrIn5 and CeIn3 in their corresponding local-moment parameter regions~\cite{Shim07,Lu16,Lu20}.
Therefore, we obtain another characteristic temperature $T_{\mathrm{KS}}\approx$ 600 K, below which the conduction $spd$ electrons start to screen the localized Ce-4$f$ electrons to form Kondo singlet states, and consequently generate Kondo resonance peak gradually in $4f$ DOS near the Fermi level.
Apart from the resonance peaks, the on-site Coulomb repulsion between 4$f$ electrons also produces broad lower and upper Hubbard bands in the 4$f$ DOS, concentrated mainly between (-3, -1) eV and (1.5, 4) eV from the Fermi level respectively, and the distance between their centers is roughly 4.3 eV, comparable with Hubbard strength $U$=5 eV. Besides, there is very little DOS weight in the lower Hubbard band thus hard to identify, while the upper Hubbard band has large intensity in the DOS.

\begin{figure}[tbp]
\hspace{-0cm} \includegraphics[totalheight=1.7in]{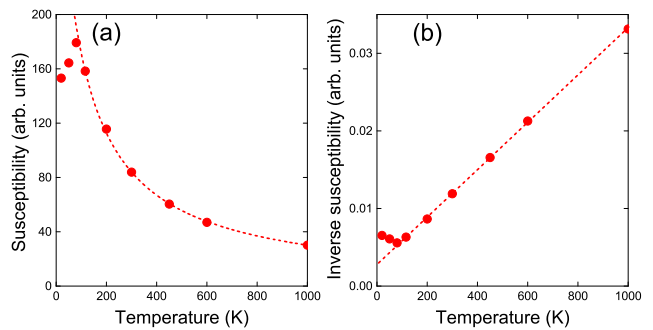}
\caption{
Red dots show the temperature dependence of (a) local magnetic susceptibility $\chi_s$  and (b) inverse susceptibility $\chi^{-1}_s$ for CePt$_2$Si$_2$, via DMFT calculations. For comparison, the Curis-Weiss form $\chi_s=C/(T+\theta)$ is denoted by red dashed lines.
}
\label{susc}
\end{figure}

We now turn to the magnetic susceptibility of CePt$_2$Si$_2$. In Fig.\ref{susc}, the local spin susceptibility $\chi_s$ and its inverse $\chi^{-1}_s$ are illustrated as functions of temperature, computed during the CT-QMC loop in DMFT iterations. In Fig.\ref{susc}(a), $\chi_s$ shows a faster increase upon cooling, then undergoes an abrupt decrease below 80 K.
At $T>$80 K, $\chi_s$ can be well fitted by the Curie-Weiss form $\chi_s=C/(T+\theta)$ with $\theta\approx$ 92 K, and starts to deviate from the Curie-Weiss formula considerably below 80 K.
Such temperature dependence of magnetic susceptibility shows consistence with experimental results~\cite{Tchokonte05,Ayache87,Tchokonte01}, and provides strong evidence of the appearance of Kondo coherence below coherence temperature $T_{\mathrm{coh}}=$ 80 K, similar to Kondo semimetal CeFe$_2$Al$_{10}$~\cite{Nam21}.

%In addition to magnetic susceptibility, the measured specific heat and resistivity
%of CePt$_2$Si$_2$ both show typical Kondo lattice behavior with their maximum value
%at 60-70 K~\cite{Ayache87,Dommann85,Tchokonte05}, and provide further evidence of
%Kondo coherence.

The above analyses of 4$f$ self-energies, Kondo resonance peak and magnetic susceptibility clearly witness a Kondo mechanism of CePt$_2$Si$_2$. Above a crossover temperature $T_{\mathrm{KS}}\approx600$ K, the Ce-4$f$ electrons are tightly bound and fully localized inside Ce atoms, thus are totally decoupled from itinerant electrons to form local moments. Reduction of temperature from $T_{\mathrm{KS}}$ induces local Kondo screening of local moments by conduction electrons gradually, results in Kondo singlet states and arouses Kondo resonance peak near the Fermi level. Further cooling to below $T_{\mathrm{coh}}\approx80$ K drives additional indirect non-local Ruderman-Kittel-Kasuya-Yosida (RKKY) interaction between $f$ electrons~\cite{Burdin00}, which creates a Kondo-coherent many-body state and gives rise to intense Kondo resonance peak and well resolved heavy-fermion hybridization bands.

\section{The Dirac node lines}

\begin{figure*}[tbp]
\hspace{-0cm} \includegraphics[totalheight=2.6in]{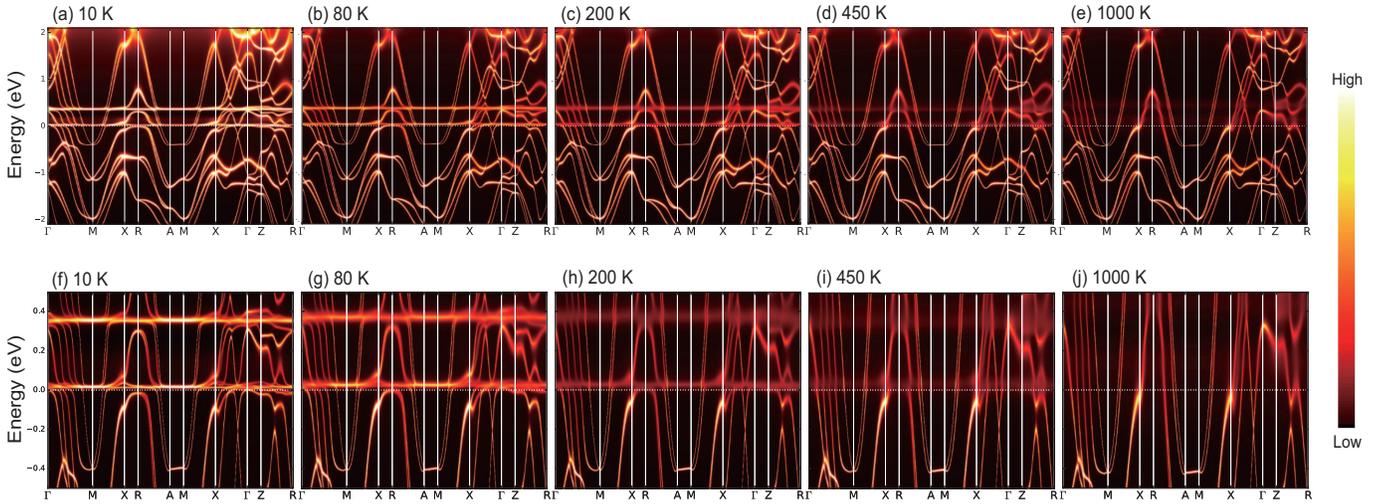}
\caption{
DFT+DMFT momentum-resolved spectral function of CePt$_2$Si$_2$ from 10 K to 1000 K. The bottom patterns are just zoomed-in view of top patterns. From 10 K to 80 K, the hybridization bands near the Fermi level are clearly resolved, while at 1000 K, the hybridization bands are already blurred out (in (e) and (i)), leaving highly-dispersive conduction bands and signaling local-moment nature of Ce-4$f$ electrons at high temperature.
}
\label{specfunc}
\end{figure*}

\begin{figure}[tbp]
\hspace{-0cm} \includegraphics[totalheight=2.55in]{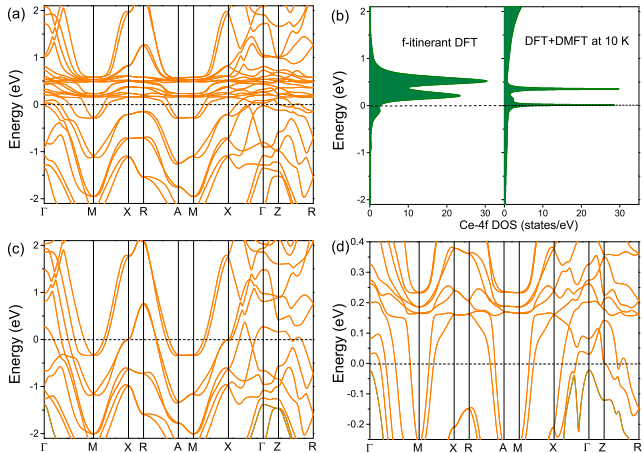}
\caption{
DFT bands of CePt$_2$Si$_2$ treating Ce-4$f$ electrons as (a) itinerant and (c) open-core. (d) is just a zoomed-in view of (a) near the Fermi level. The bands are degenerate at X, R, M, A and along X-R and M-A paths, leading to DNLs locating on X-R and M-A lines shown in Fig.\ref{lattice}(c).
Ce-4$f$ bands in (a) has been divided into $4f_{5/2}$ and $4f_{7/2}$ bands under SOC splitting, contributing double 4$f$ DOS peaks in (b), by comparison, DFT+DMFT calculation leads to much narrow 4$f$ DOS peaks at 10 K.}
\label{DFTbands-fdos}
\end{figure}

\begin{figure*}[tbp]
\hspace{0cm} \includegraphics[totalheight=2.2in]{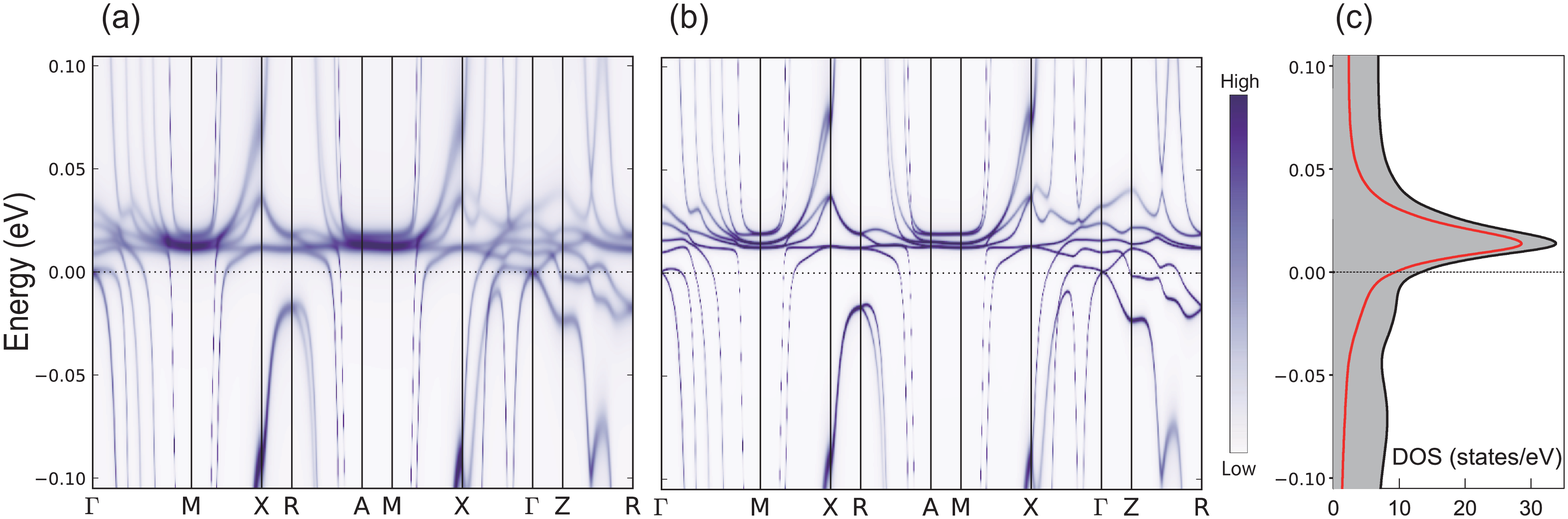}
\caption{
(a) DFT+DMFT momentum-resolved spectral function of CePt$_2$Si$_2$ near Fermi energy at 10 K. (b) Modify $\mathrm{Im}\Sigma_{5/2}(\omega)$ to see the low-lying hybridization bands and their crossings more clearly. (c) DFT+DMFT DOS at 10 K, in which the black line denotes total DOS, red line shows
DOS of $4f_{5/2}$ state.
}
\label{10Klimit}
\end{figure*}

The transition to Kondo coherent state below $T_{\mathrm{coh}}$ can be witnessed by temperature variation of spectral function, which can be measured directly via ARPES experiments. In Fig.\ref{specfunc}, the momentum-resolved spectral function $A(\mathbf{k},\omega)$ of CePt$_2$Si$_2$ calculated by DFT+DMFT is plotted along a representative high-symmetry path in the Brillouin zone, in temperature range from 10 K to 1000 K. At low temperatures (10 K and 80 K cases), the intense spectral weight of two groups of weakly dispersive heavy-fermion bands, which concentrate near 14 meV and 0.351 eV above the Fermi level, respectively, can be clearly seen, through which the two narrow resonance peaks in DOS (Fig.\ref{DOS}(a)) can be directly obtained via momentum integral of spectral function through $\rho(\omega)=\sum_{\mathbf{k}}A(\mathbf{k},\omega)$. These two groups of hybridization bands are constructed through $c$-$f$ hybridization between conduction $spd$ electrons with Ce-4$f_{5/2}$ or 4$f_{7/2}$ states, respectively, resulting in heavy-fermion quasi-particles with enhanced effective mass $m^{\ast}$ determined by Eq.\ref{Z}. The sharply resolved heavy-fermion bands near the Fermi level below 80 K again verifies formation of Kondo coherence below $T_{\mathrm{coh}}\approx$80 K. As temperature rises from $T_{\mathrm{coh}}$, the hybridization bands become more and more blurred and are no longer well resolved.

Above the local Kondo screening temperature $T_{\mathrm{KS}}\approx$ 600 K, the intensity of hybridization bands is dramatically diminished (see Fig.\ref{specfunc}(e) at 1000K), leading to small magnitude of 4$f$ DOS peaks (see the top pattern in Fig.\ref{DOS}(a)), consequently the 4$f$ electrons are fully localized, and the bands near the Fermi level become highly dispersive, see Fig.\ref{specfunc}(j). Fig.\ref{DFTbands-fdos}(c) shows the DFT bands treating 4$f$ orbits as open-core states~\cite{Chen21}, which can nicely reproduce high-temperature DFT+DMFT result in Fig.\ref{specfunc}(e), further confirming the local-moment nature of 4$f$ states at $T>T_{\mathrm{KS}}$.
It should be noted that the spectral weight of $4f$ electrons remains non-vanished even at $T>T_{\mathrm{KS}}$, similar to $\gamma$-Ce, CeSb and CeIn$_3$ in their local-moment regions~\cite{Lu16,Lu20,Zhu20}.

At 10 K, the DFT+DMFT hybridization bands around 14 meV above the Fermi level are clearly distinguishable, see Fig.\ref{specfunc}(f) and enlarged view in Fig.\ref{10Klimit}(a), in which the band structure is indeed similar to that of the DFT bands in Fig.\ref{DFTbands-fdos}(a) and (d) which treat Ce-4$f$ electrons to be itinerant. The hybridization bands of $f$-itinerant DFT also split into $4f_{5/2}$ and $4f_{7/2}$ bands by an energy interval 0.31 eV, close to the SOC splitting by 0.337 eV in DFT+DMFT simulation. As shown in Fig.\ref{DFTbands-fdos}(b), the 4$f$ DOS shows three major differences between DFT and DFT+DMFT results, firstly, the 4$f$ peak width (198 meV in DFT) is strongly reduced to 20 meV in DFT+DMFT under electron correlations, with a reduction factor of 9.9 close to the mass enhancement $m^{\ast}/m_{\mathrm{DFT}}=1/Z=10.3$; secondly, in DFT+DMFT result, the $4f_{5/2}$ DOS shifts towards the Fermi level, forming Kondo resonance peak very close to the Fermi energy, while the DFT $4f_{5/2}$ DOS peak locates considerable higher above the Fermi level; thirdly, DFT+DMFT produces additional lower and upper Hubbard bands far away from the Fermi level.

In the literature, it has been shown that the nonsymmorphic symmetries combining point group and fractional translation operations can generate additional degeneracy along certain high-symmetry paths in the Brillouin zone, which are robust under action of SOC~\cite{Young15}. In space group No. 129, the nonsymmorphic symmetries are the gliding mirror plane $\{M_z|\frac{1}{2},\frac{1}{2}\}$ and screw axes $\{C_{2x}|\frac{1}{2},0\}$, $\{C_{2y}|0,\frac{1}{2}\}$, which are hold by CePt$_2$Si$_2$ and CePt$_2$Ge$_2$ crystals in Fig.\ref{lattice}(b). At $T>T_{\mathrm{KS}}$, since Ce-4$f$ electrons are already localized, the electron bands of CePt$_2$Si$_2$ can be reflected by 4$f$ open-core DFT results in Fig.\ref{DFTbands-fdos}(c). With time-reversal symmetry in their paramagnetic phases, the space inversion symmetry of CePt$_2$Si$_2$ and CePt$_2$Ge$_2$ guarantees global two-fold degeneracy of the electron bands, and the nonsymmorphic symmetries give additional degeneracy of bands at X, R, M, A points, therefore, four-fold Dirac crossings arise, generating Dirac nodes at these points.
Moreover, degeneracy of bands remain along X-R and M-A paths (see Fig.\ref{DFTbands-fdos}(c)), resulting in DNLs, similar to ZrSiS and PtPb4~\cite{Schoop15,Fu17,Chen17,Wu22}.
The X-R and M-A DNLs locate at the boundary of the Brillouin zone, as shown in Fig.\ref{lattice}(c). The energy ranges of the X-R DNLs in CePt$_2$Si$_2$ are (-0.725 eV, -0.624 eV) and (6.4 meV, 0.759 eV) from the Fermi level, while for M-A DNLs it is in (-0.338 eV, -0.330 eV).
At $T>T_{\mathrm{KS}}$, the similarity between spectral function (Fig.\ref{specfunc}(e)) and 4$f$ open-core bands in Fig.\ref{DFTbands-fdos}(c) clearly verifies above analysis of band-crossings and appearance of DNLs in CePt$_2$Si$_2$ in local-moment region. It should be stressed that since the Ce-4$f$ electrons are already localized at $T>T_{\mathrm{KS}}$, such DNLs are non-interacting and are composed by light conduction electrons (mostly Ce-$d$, Pt-$p$,$d$, and Si-$p$ electrons).

We have shown in Fig.\ref{specfunc}(f) and Fig.\ref{10Klimit}(a) that below $T_{\mathrm{coh}}\approx$ 80 K, the hybridization bands can be clearly identified near the Fermi level, in which their crossings at X, R and along X-R path are already legible, while at M, A and along M-A path seem a little fuzzy because several crossings concentrate in a narrow energy range. In order to see the DNLs more clearly, we slightly reduce the value of $\Sigma_0$ in the parabolic expression of imaginary self-energy $\mathrm{Im}\Sigma_{5/2}(\omega)$ and recalculate the spectral. The obtained spectral function is displayed in Fig.\ref{10Klimit}(b), in which the locations of crossings between hybridization bands match $f$-itinerant DFT results in Fig.\ref{DFTbands-fdos}(d), confirming the existence of interacting DNLs along X-R and M-A paths, since such crossings are protected by lattice nonsymmorphic symmetries and robust under electron correlations.
Nevertheless, below $T_{\mathrm{coh}}$, electron correlations push the hybridization bands much closer to the Fermi level than $f$-itinerant DFT results, hence generate much narrow energy windows for X-R and M-A DNLs, which are about (-90 meV, -15 meV) and (12.5 meV, 37 meV) along X-R, and (12.5 meV, 18.8 meV) along M-A, from the Fermi level. Compare Fig.\ref{10Klimit}(a) and (c), it can be seen that the energy window of M-A DNLs locates at the center of the Kondo resonance peak, therefore, the Dirac fermions near M-A DNLs are highly renormalized and are essentially the heavy-fermions hybridizing $4f_{5/2}$ electrons with conduction electrons.
The energy window of X-R DNLs locates at the lower tail of the Kondo resonance peak, thus the Dirac fermions near X-R DNLs are also interacting.

\begin{figure*}[tbp]
\hspace{0cm} \includegraphics[totalheight=2.75in]{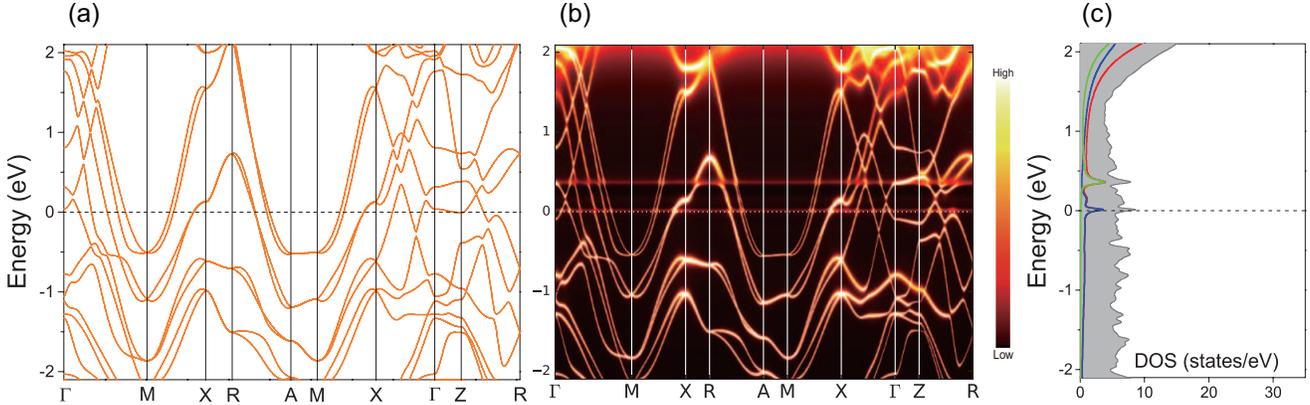}
\caption{
(a) 4f Open-core DFT bands of CePt$_2$Ge$_2$ and (b) the momentum-resolved spectral function at 10 K via DFT+DMFT calculation. (c) Projected DOS by DFT+DMFT simulation at 10 K, in which the dashed horizontal line represents the Fermi level, grey solid line denotes total DOS, red line denotes Ce-4f DOS, blue and green lines show projected Ce-$4f_{5/2}$ and $4f_{7/2}$ DOS, respectively.
}
\label{Gebands}
\end{figure*}

From above discussions, we can now verify the appearance of interacting DNLs in CePt$_2$Si$_2$, locating along X-R and M-A paths in its Brillouin zone, driven by Kondo coherence below coherent temperature $T_{\mathrm{coh}}\approx$ 80 K, and the energy windows of these DNLs are very close to the Fermi level. The Dirac fermions of these interacting DNLs are constructed by heavy-fermions with strongly enhanced effective mass and reduced velocity.
As temperature rises from $T_{\mathrm{coh}}$, the interacting DNLs are destructed gradually and are no longer clearly identified. Above the local Kondo screening temperature $T_{\mathrm{KS}}\approx$ 600 K, the DNLs reappear near the Fermi level, but now the correlated 4$f$ electrons are already localized, so now the Dirac fermions are composed of non-correlated light conduction electrons.
Besides, the energy ranges of the DNLs in these two cases are shifted dramatically, in that the energy of the non-interacting DNLs is much far from the Fermi level than the interacting DNLs.

Now we turn to discuss CePt$_2$Ge$_2$.
By contrast, the transition to Kondo-coherent state upon cooling in CePt$_2$Si$_2$ does not emerge in CePt$_2$Ge$_2$. In Fig.\ref{Gebands}, we compare the 4$f$ open-core DFT bands with the DFT+DMFT momentum-resolved spectral function at 10 K for CePt$_2$Ge$_2$, which shows good correspondence. Even at such low temperature, the Ce-4$f$ states just contribute a small DOS near the Fermi level, and the 4$f$ peak height is much smaller than that of CePt$_2$Si$_2$ at low temperature. Besides, at 10 K, the imaginary $4f_{5/2}$ self-energy of CePt$_2$Ge$_2$ has very large value at $\omega=0$, similar to CeSb and $\gamma$-Ce which are in local-moment region.
Therefore, the 4$f$ electrons are localized, suggesting the local-moment nature of Ce-4$f$ states in CePt$_2$Ge$_2$, which can be directly verified by the similarity of 4$f$ open-core DFT bands with the spectral function in Fig.\ref{Gebands}. The strong suppression of Ce-4$f$ resonance peak in CePt$_2$Ge$_2$ results from about $7\%$ volume increase of the unit cell than CePt$_2$Si$_2$, which also drives the itinerant-localized shift of $f$ electrons in CeIn$_3$ and $\alpha-\gamma$ transition of Ce metal~\cite{Chen18,Lu16}. Since the correlated 4$f$ electrons are localized, the X-R and M-A DNLs in CePt$_2$Ge$_2$ are non-interacting, and their energy ranges are (-0.708 eV, -0.615 eV) and (0.129 eV, 0.731 eV) for X-R, (-0.521 eV, -0.507 eV) for M-A, all relatively far from the Fermi level.

\section{conclusion and discussion}

To summarize, we have performed systematic DFT+DMFT simulations of CePt$_2$Si$_2$ and CePt$_2$Ge$_2$ in a wide temperature range. By examining the temperature dependence of Ce-4$f$ self-energies, Kondo resonance peak, magnetic susceptibility, and momentum-resolved spectral for CePt$_2$Si$_2$, we have verified a crossover from localization of Ce-4$f$ electrons to local Kondo screening of 4$f$ electrons by $spd$ conduction electrons, at Kondo screening temperature $T_{\mathrm{KS}}\approx$ 600 K upon cooling. Secondly, as temperature decreases further, a transition takes place from local Kondo screening to Kondo coherent state in CePt$_2$Si$_2$ at coherence temperature $T_{\mathrm{coh}}\approx$ 80 K, and the Kondo coherence is driven by indirect RKKY interaction between 4$f$ electrons.
In contrast, due to larger unit-cell volume than CePt$_2$Si$_2$, Ce-4$f$ electrons in CePt$_2$Ge$_2$ remain localized as low as 10 K.

The lattice nonsymmorphic symmetries in CePt$_2$Si$_2$ and CePt$_2$Ge$_2$ give rise to symmetry-protected DNLs along X-R and M-A high-symmetry paths in the Brillouin zone, which are robust under action of SOC and electron correlations.
For CePt$_2$Si$_2$ above $T_{\mathrm{KS}}$, the local-moment nature of Ce-4$f$ electrons makes these DNLs non-interacting, mainly composing by light $spd$ conduction electrons, and the energy windows of these DNLs are relatively far from the Fermi level. Below $T_{\mathrm{coh}}$, the emergence of Kondo coherence drives the DNLs in CePt$_2$Si$_2$ strongly interacting and constructing by heavy-fermions with strongly enhanced effective mass and reduced velocity. Remarkably, the energy range of the interacting DNLs is shifted to be much closer to the Fermi level.
In intermediate temperature range $T_{\mathrm{coh}}<T<T_{\mathrm{KS}}$, the DNLs are no longer well identified.
By comparison, in CePt$_2$Ge$_2$, the DNLs along X-R and M-A paths are non-interacting in wide temperature region, consisting of only light conduction electrons.

It is well known that in some rare-earth semimetallic compounds, Weyl fermions can arise due to breaking of time-reversal symmetry by magnetic order, such as in Weyl node-point semimetals CeSb~\cite{Guo17,Fang20,Lu20}, CeBi~\cite{Huan21,Matt22}, CeAlGe, CeAlSi and LaAlSi~\cite{Chang18,Puphal20,Su21,Sakhya22}; or arise by breaking of space-inversion symmetry in noncentrosymmetric lattices, such as in Weyl node-point semimetal CeRu$_4$Sn$_6$~\cite{Xu17}, and  Weyl node-ring semimetals Ce$_3$Bi$_4$Pt$_3$~\cite{Lai18,Cao20} and YbCdGe~\cite{Laha19}.
In addition, Dirac fermions also emerge in node-line Dirac semimetals CeRhSb and CeNiSn with centrosymmetric structures~\cite{Nam19}. Although model studies have implied that heavy-Weyl or heavy-Dirac quasiparticles can emerge in Anderson lattice model~\cite{Chang18-2,Grefe20}, their appearance in real materials are rare. In most of above materials (except for Weyl semimetals CeRu$_4$Sn$_6$ and Ce$_3$Bi$_4$Pt$_3$), the $f$ electrons in rare-earth atoms are actually localized, so the Weyl or Dirac fermions in these materials are composed of non-interacting conduction electrons. It should be also stressed that nonsymmorphic CeSbTe also holds DNLs in its paramagnetic phase, however, CeSbTe shows a magnetic order below 2.7 K, and the Ce-4$f$ states are essentially localized in its paramagnetic phase~\cite{Schoop18}, so their DNLs are formed by conduction electrons in all temperature region.
In this context, our revealed interacting DNLs in CePt$_2$Si$_2$ provides a very rare case of interacting Dirac fermions in real materials, besides, the dramatic change of Dirac fermions near the DNLs at different temperature regions provides an unique platform for future experimental investigations.

The bulk DNLs and induced surface states have been observed by ARPES experiments in nonsymmorphic ZrSiS~\cite{Chen17,Schoop15,Fu17} and PtPb4~\cite{Wu22}, which exhibit non-interacting DNLs on the boundaries of their Brillouin zones,
likewise, the interacting DNLs we proposed in CePt$_2$Si$_2$ also locate along high-symmetric X-R and M-A lines, making it easier to be observed experimentally, and the drastic temperature variation of dispersions near the DNLs can also be detected conveniently. We find that the surface states induced by DNLs in CePt$_2$Si$_2$ depend sensitively on the cleavage plane of the crystal, hence it requires further experimental data to obtain detailed surface dispersions. Besides, the interacting DNLs below $T_{\mathrm{coh}}$ may give rise to heavy surface states which are distinct to ordinary light surface states above $T_{\mathrm{KS}}$~\cite{Peters16}.

According to model studies of Weyl Kondo semimetals, the interacting node point or node line can give rise to unusual transport phenomena in comparison with non-interacting cases, e.g., the strong mass enhancement and velocity reduction of Weyl nodes produce a $T^3$ temperature dependence of specific heat~\cite{Lai18}, while the Weyl node line provides two-dimensional Weyl fermions in momentum space, hence contributes a $T^2$ dependence of specific heat~\cite{Chen21}. Moreover, the electron correlations in topological semimetals can induce nonlinear-response phenomena such as giant spontaneous Hall effect, as observed in node-line Weyl semimetal Ce$_3$Bi$_4$Pt$_3$~\cite{Dzsaber21,Kofuji21}. Similarly, the interacting DNLs in CePt$_2$Si$_2$ may also induced anomalous nonlinear responses, which deserves further experimental and theoretical investigations.

\acknowledgments
This work is supported by GuikeAD20159009, National Natural Science Foundation of China (NO. 12004048, 11864008 and 11764010), the National Key Research and Development Program of China (No. 2021YFB3501503), and the Foundation of LCP.


\begin{thebibliography}{99}

\bibitem{Wang12} Zhijun Wang, Yan Sun, Xing-Qiu Chen, Cesare Franchini, Gang Xu, Hongming Weng, Xi Dai, and Zhong Fang, Phys. Rev. B \textbf{85}, 195320 (2012).
%Dirac semimetal and topological phase transitions in A3Bi (A = Na, K, Rb)

\bibitem{Wang13} Zhijun Wang, Hongming Weng, Quansheng Wu, Xi Dai, and Zhong Fang, Phys. Rev. B \textbf{88}, 125427 (2013).
%Three-dimensional Dirac semimetal and quantum transport in Cd3As2,

\bibitem{Lv15} B. Q. Lv, H. M. Weng, B. B. Fu, X. P. Wang, H. Miao, J. Ma, P. Richard, X. C. Huang, L. X. Zhao, G. F. Chen, Z. Fang, X. Dai, T. Qian, and H. Ding, Phys. Rev. X \textbf{5}, 031013 (2015).
%Experimental Discovery of Weyl Semimetal TaAs,

\bibitem{Wang19} Zhenwei Wang, Kaifa Luo, Jianzhou Zhao, Rui Yu, , Phys. Rev. B \textbf{100}, 205117 (2019).
%Large Fermi arc and robust Weyl semimetal phase in Ag2S

\bibitem{Fu17} B.-B. Fu, C.-J. Yi, T.-T. Zhang, M. Caputo, J.-Z. Ma, X. Gao, B. Q.
Lv, L.-Y. Kong, Y.-B. Huang, M. Shi, V. N. Strocov, C. Fang, H.-M.
Weng, Y.-G. Shi, T. Qian, and H. Ding,  arXiv:1712.00782.
%Observation of bulk nodal lines in topological semimetal ZrSiS

\bibitem{Schoop15} Leslie M. Schoop, Mazhar N. Ali, Carola Stra{\ss}er, Andreas Topp, Andrei Varykhalov, Dmitry Marchenko,
Viola Duppel, Stuart S. P. Parkin, Bettina V. Lotsch, and Christian R. Ast,
Nature Communications \textbf{7}, 11696 (2015).
%Dirac cone protected by non-symmorphic symmetry and three-dimensional Dirac line node in ZrSiS

\bibitem{Chen17} C. Chen, X. Xu, J. Jiang, S.-C. Wu, Y. P. Qi, L. X. Yang, M. X. Wang, Y. Sun, N. B. M. Schr\"{o}ter,
H. F. Yang, L. M. Schoop, Y. Y. Lv, J. Zhou, Y. B. Chen, S. H. Yao, M. H. Lu,
Y. F. Chen, C. Felser, B. H. Yan, Z. K. Liu, and Y. L. Chen, Phys. Rev. B \textbf{95}, 125126 (2017).
%Dirac line nodes and effect of spin-orbit coupling in the nonsymmorphic critical semimetals MSiS (M = Hf, Zr)

\bibitem{Wu22} Han Wu, Alannah M. Hallas, Xiaochan Cai, Jianwei Huang, Ji Seop Oh, Vaideesh Loganathan, Ashley Weiland, Gregory T. McCandless, Julia Y. Chan,
Sung-Kwan Mo, Donghui Lu, Makoto Hashimoto, Jonathan Denlinger, Robert J. Birgeneau,
Andriy H. Nevidomskyy, Gang Li, Emilia Morosan, and Ming Yi, npj Quantum Materials \textbf{7}, 31 (2022).
%Nonsymmorphic symmetry-protected band crossings in a square-net metal PtPb4

\bibitem{Young15} Steve M. Young, Charles L. Kane, Phys. Rev. Lett. \textbf{115}, 126803 (2015).
%Dirac Semimetals in Two Dimensions

\bibitem{Lai18} Hsin-Hua Lai, Sarah E. Grefe, Silke Paschen, and Qimiao Si, Proc. Natl. Acad. Sci. USA \textbf{115}, 93 (2018).
%Weyl¨CKondo semimetal in heavy-fermion systems

\bibitem{Cao20}
Chao Cao, Guo-Xiang Zhi, and Jian-Xin Zhu, Phys. Rev. Lett. \textbf{124}, 166403 (2020).
%From Trivial Kondo Insulator Ce3Pt3Bi4 to Topological Nodal-Line Semimetal Ce3Pd3Bi4

\bibitem{Kofuji21} Akira Kofuji, Yoshihiro Michishita, and Robert Peters, Phys. Rev. B \textbf{104}, 085151 (2021).
%Efects of strong correlations on the nonlinear response in Weyl-Kondo semimetals

\bibitem{Dzsaber21}
S. Dzsaber, X. Yan, M. Taupin, G. Eguchi, A. Prokofiev, T. Shiroka, P. Blaha, O. Rubel, S. E. Grefe, H.-H. Lai, Q. Si, and S. Paschen, Proc. Natl. Acad. Sci. USA \textbf{118}, e2013386118 (2021).
%Giant spontaneous Hall effect in a nonmagnetic Weyl-Kondo semimetal,

\bibitem{Nam19} T.-S. Nam, Chang-Jong Kang, D.-C. Ryu, Junwon Kim, Heejung Kim, Kyoo Kim, and B. I. Min, Phys. Rev. B \textbf{99}, 125115 (2019).
%Topological bulk band structures of the hourglass and Dirac nodal-loop types in Ce Kondo systems: CeNiSn, CeRhAs, and CeRhSb

\bibitem{Dommann85} A. Dommann, F. Hulliger, and H. R. Ott, V. Gramlich,
Journal of the Less-Common Metals \textbf{110}, 331 (1985).
%THE CRYSTAL STRUCTURE AND SOME PROPERTIES OF CePt2Si2 AND CePt2Ge2

\bibitem{Reotier97} P. Dalmas de R\'{e}otier, A. Yaouanc, R. Calemczuk, A. D. Huxley, C. Marcenat, P. Bonville, P. Lejay, P. C. M. Gubbens, and A. M. Mulders, Phys. Rev. B \textbf{55}, 2737 (1997).
%CePt 2Si 2: A Kondo lattice compound with no magnetic ordering down to 0.06 K

\bibitem{Haule10} K. Haule, C.-H. Yee, and K. Kim, Phys. Rev.
B \textbf{81}, 195107 (2010).
%Dynamical meanfield theory within the full-potential methods: Electronic structure of CeIrIn5, CeCoIn5, and CeRhIn5,

\bibitem{Blaha20} P. Blaha, K. Schwarz, F. Tran, R. Laskowski, G. K. H. Madsen, and L. D. Marks, J. Chem. Phys. \textbf{152}, 074101 (2020).
%WIEN2k: An APW+lo program for calculating the properties of solids,

\bibitem{Lu16}
Haiyan Lu and Li Huang, Phys. Rev. B \textbf{94}, 075132 (2016).
%Pressure-driven 4 f localized-itinerant crossover in heavy-fermion compound CeIn3: A first-principles many-body perspective

\bibitem{Shim07} J. H. Shim, K. Haule, and G. Kotliar, Science \textbf{318}, 1615 (2007).
%Modeling the Localized-to-ItinerantElectronic Transition in the HeavyFermion System CeIrIn5

\bibitem{Zhu20}
Xie-Gang Zhu, Yu Liu, Ya-Wen Zhao, Yue-Chao Wang, Yun Zhang, Chao Lu, Yu Duan, Dong-Hua Xie, Wei Feng, Dan Jian,
Yong-Huan Wang, Shi-Yong Tan, Qin Liu, Wen Zhang, Yi Liu, Li-Zhu Luo, Xue-Bing Luo, Qiu-Yun Chen,
Hai-Feng Song, and Xin-Chun Lai, npj Quantum Materials \textbf{5}, 47 (2020).
%Kondo scenario of the ¦Ã¨C¦Á phase transition in single crystalline cerium thin films

\bibitem{Wang21} Yue-Chao Wang, Yuan-Ji Xu, Yu Liu, Xing-Jie Han, Xie-Gang Zhu, Yi-feng Yang, Yan Bi, Hai-Feng Liu, and Hai-Feng Song, Phys. Rev. B \textbf{103}, 165140 (2021).
%First-principles study of the role of surface in the heavy-fermion compound CeRh2Si2

\bibitem{Chen18}
Q. Y. Chen, W. Feng, D. H. Xie, X. C. Lai, X. G. Zhu, and L. Huang, Phys. Rev. B \textbf{97}, 155155 (2018).
%Localized to itinerant transition of f electrons in ordered Ce films on W(110)

\bibitem{Ayache87} C. Ayache, J. Beille, E. Bonjour, R. Calemczuk, G. Creuzet,
D. Gignoux, A. Najib, D. Schmitt, J. Voiron, and M. Zerguine, Journal of Magnetism and Magnetic Materials \textbf{63, 64}, 329 (1987).
%SPECIFIC HEAT AND PRESSURE EFFECTS IN THE KONDO LATTICE COMPOUND CePt2Si2

\bibitem{Tchokonte05} M.B. Tchoula Tchokont\'{e}, P. de V. du Plessisb, A.M. Strydom, Solid State Communications \textbf{136}, 450 (2005).
%Kondo lattice behaviour in CePt2(Si1KxSnx)2 alloys

\bibitem{Lu20}
Haiyan Lu and Qin Liu, J. Phys.: Condens. Matter \textbf{32}, 485601 (2020).
%Exploring the exotic f states of prototype compounds CeSb and USb

\bibitem{Tchokonte01} M.B. Tchoula Tchokont\'{e}, P. de V. du Plessis, A.M. Strydom, D. Kaczorowski, Journal of Magnetism and Magnetic Materials \textbf{226}, 173 (2001).
%Magnetic and electrical properties of the Kondo system CePt(SiGe)

\bibitem{Nam21} T.-S. Nam, Junwon Kim, Chang-Jong Kang, Kyoo Kim, and B. I. Min, Phys. Rev. B \textbf{103}, 045101 (2021).
%Temperature-dependent electronic structure and topological property of the Kondo semimetal CeFe2Al10

\bibitem{Burdin00} S. Burdin, A. Georges, D. R. Grempel, Phys. Rev. Lett. \textbf{85}, 1048 (2000).
%Coherence scale of the Kondo lattice

\bibitem{Chen21} Lei Chen, Chandan Setty, Haoyu Hu, Maia G. Vergniory, Sarah E. Grefe, Andrey Prokofiev,
Silke Paschen, Jennifer Cano, and Qimiao Si, arXiv: 2107.10837.
%Topological Semimetal Driven by Strong Correlations and Crystalline Symmetry

\bibitem{Guo17}
Chunyu Guo, Chao Cao, Michael Smidman, Fan Wu, Yongjun Zhang, Frank Steglich, Fu-Chun Zhang, and Huiqiu Yuan, npj Quantum Materials \textbf{2}, 39 (2017).
%Possible Weyl fermions in the magnetic Kondo system CeSb

\bibitem{Fang20}
Y. Fang, F. Tang, Y. R. Ruan, J. M. Zhang, H. Zhang, H. Gu, W. Y. Zhao, Z. D. Han, W. Tian, B. Qian,
X. F. Jiang, X. M. Zhang, and X. Ke, Phys. Rev. B \textbf{101}, 094424 (2020).
%Magnetic-field-induced nontrivial electronic state in the Kondo-lattice semimetal CeSb

\bibitem{Huan21}
Shuchun Huan, Xianbiao Shi, Lixuesong Han, Hao Su, Xia Wang, Zhiqiang Zou,
Na Yua, Weiwei Zhao, Leiming Chene, Yanfeng Guo, Journal of Alloys and Compounds \textbf{875}, 159993 (2021).
%Magnetotransport evidence for the nontrivial topological states in the fully spin-polarized Kondo semimetal CeBi

\bibitem{Matt22}
Christian E. Matt, Yu Liu, Harris Pirie, Nathan C. Drucker, Na Hyun Jo, Brinda Kuthanazhi,
Zhao Huang, Christopher Lane, Jian-Xin Zhu, Paul C. Canfield, and Jennifer E. Hoffman, Phys. Rev. B \textbf{105}, 085134 (2022).
 %Spin-polarized imaging of strongly interacting fermions in the ferrimagnetic state of Weyl candidate CeBi

\bibitem{Chang18} Guoqing Chang, Bahadur Singh, Su-Yang Xu, Guang Bian, Shin-Ming Huang, Chuang-Han Hsu,
Ilya Belopolski, Nasser Alidoust, Daniel S. Sanchez, Hao Zheng, Hong Lu, Xiao Zhang, Yi Bian, Tay-Rong Chang,
Horng-Tay Jeng, Arun Bansil, Han Hsu, Shuang Jia, Titus Neupert, Hsin Lin,,and M. Zahid Hasan, Phys. Rev. B \textbf{97}, 041104(R) (2018).
%Magnetic and noncentrosymmetricWeyl fermion semimetals in the RAlGe family of compounds (R = rare earth)

\bibitem{Puphal20} Pascal Puphal, Vladimir Pomjakushin, Naoya Kanazawa, Victor Ukleev, Dariusz J. Gawryluk,
Junzhang Ma, Muntaser Naamneh, Nicholas C. Plumb, Lukas Keller, Robert Cubitt,
Ekaterina Pomjakushina, and Jonathan S. White, Phys. Rev. Lett. \textbf{124}, 017202 (2020).
%Topological Magnetic Phase in the Candidate Weyl Semimetal CeAlGe

\bibitem{Su21} Hao Su, Xianbiao Shi, Jian Yuan, Yimin Wan, Erjian Cheng, Chuanying Xi, Li Pi, Xia Wang, Zhiqiang Zou,
Na Yu, Weiwei Zhao, Shiyan Li, and Yanfeng Guo, Phys. Rev. B \textbf{103}, 165128 (2021).
%Multiple Weyl fermions in the noncentrosymmetric semimetal LaAlSi

\bibitem{Sakhya22} Anup Pradhan Sakhya, Cheng-Yi Huang, Gyanendra Dhakal, Xue-Jian Gao,
Sabin Regmi, Xiaohan Yao, Robert Smith, Milo Sprague, Bahadur Singh,
Hsin Lin, Su-Yang Xu, Fazel Tafti, Arun Bansil, and Madhab Neupane, arXiv:2203.05440.
%Observation of Fermi arcs and Weyl nodes in a non-centrosymmetric magnetic Weyl semimetal


\bibitem{Xu17} Yuanfeng Xu, Changming Yue, Hongming Weng, and Xi Dai, Phys. Rev. X \textbf{7}, 011027 (2017).
%Heavy Weyl Fermion State in CeRu4Sn6

\bibitem{Laha19} Antu Laha, Sudip Malick, Ratnadwip Singha, Prabhat Mandal, P. Rambabu, V. Kanchana, and Z. Hossain, Phys. Rev. B \textbf{99}, 241102(R) (2019).
%Magnetotransport properties of the correlated topological nodal-line semimetal YbCdGe

\bibitem{Grefe20} Sarah E. Grefe, Hsin-Hua Lai, Silke Paschen, and Qimiao Si, Phys. Rev. B \textbf{101}, 075138 (2020).
    %Weyl-Kondo semimetals in nonsymmorphic systems
\bibitem{Chang18-2} Po-Yao Chang and Piers Coleman, Phys. Rev. B \textbf{97}, 155134 (2018).
%Parity-violating hybridization in heavyWeyl semimetals

\bibitem{Schoop18} Leslie M. Schoop, Andreas Topp, Judith Lippmann, Fabio Orlandi, Lukas M\"{u}chler, Maia G. Vergniory, Yan Sun, Andreas W. Rost, Viola Duppel, Maxim Krivenkov, Shweta Sheoran, Pascal Manuel, Andrei Varykhalov, Binghai Yan, Reinhard K. Kremer, Christian R. Ast, and Bettina V. Lotsch, Sci. Adv. \textbf{4}, 2317 (2018).
%Tunable Weyl and Dirac states in the nonsymmorphic compound CeSbTe

\bibitem{Peters16} Robert Peters, Tsuneya Yoshida, Hirofumi Sakakibara, and Norio Kawakami, Phys. Rev. B \textbf{93}, 235159 (2016).
%Coexistence of light and heavy surface states in a topological multiband Kondo insulator




\end{thebibliography}
\end{document}